\documentclass[twoside,leqno,twocolumn]{article}
\usepackage{ltexpprt}

\usepackage{calc,amsfonts,amssymb,amsmath,bm,url,color,cite}
\usepackage{graphicx,wrapfig}
\usepackage{psfrag,float, subcaption,algorithmic}

\usepackage{multirow}
\usepackage{epstopdf,booktabs,stmaryrd}
\usepackage{color}
\usepackage[dvipsnames]{xcolor}
\usepackage{pifont}

\usepackage{algorithm, epstopdf,nicefrac}

\DeclareMathOperator*{\minimize}{min}

\newcommand{\A}{\boldsymbol{A}}
\newcommand{\B}{\boldsymbol{B}}
\newcommand{\C}{\boldsymbol{C}}

\newcommand{\G}{\boldsymbol{G}}

\newcommand{\M}{\boldsymbol{M}}

\newtheorem{Prop}{Proposition}
\newtheorem{Theorem}{Theorem}

\definecolor{orange}{RGB}{255,107,0}

\begin{document}

%\setcounter{chapter}{2} % If you are doing your chapter as chapter one,
%\setcounter{section}{3} % comment these two lines out.

%\title{\Large TeKGraph: Tensor Knowledge-graph embeddings for COVID-19 drug repurposing}
\title{\Large TeX-Graph: Coupled tensor-matrix knowledge-graph embedding for COVID-19 drug repurposing}
%\author{Bo Yang \quad Ahmed Zamzam  \quad  Nicholas D. Sidiropoulos \\
%Electrical and Computer Engineering Department \\
%University of Minnesota \\
%\{yang4173, zamza002, nikos\}@umn.edu}
\author{Charilaos I. Kanatsoulis \thanks{Electrical and Computer Engineering Department, University of Minnesota. Email: kanat003@umn.edu, $ \dagger $
	 Electrical and Computer Engineering Department, University of Virginia. Email: nikos@virginia.edu}
 \footnotemark[0]
\and Nicholas D. Sidiropoulos \footnotemark[2]
} 
%	\{yang4173, zamza002, nikos\}@umn.edu}
\date{}

\maketitle

% Copyright Statement
% When submitting your final paper to a SIAM proceedings, it is requested that you include 
% the appropriate copyright in the footer of the paper.  The copyright added should be 
% consistent with the copyright selected on the copyright form submitted with the paper.
% Please note that "20XX" should be changed to the year of the meeting.

% Default Copyright Statement
% \fancyfoot[R]{\footnotesize{\textbf{Copyright \textcopyright\ 2021 by SIAM\\
% Unauthorized reproduction of this article is prohibited}}}

% Depending on which copyright you agree to when you sign the copyright form, the copyright 
% can be changed to one of the following after commenting out the default copyright statement
% above.

%\fancyfoot[R]{\footnotesize{\textbf{Copyright \textcopyright\ 20XX\\
%Copyright for this paper is retained by authors}}}

%\fancyfoot[R]{\footnotesize{\textbf{Copyright \textcopyright\ 20XX\\
%Copyright retained by principal author's organization}}}

%\pagenumbering{arabic}
%\setcounter{page}{1}%Leave this line commented out.

\begin{abstract} 
% \small\baselineskip=10pt 
Knowledge graphs (KGs) are powerful tools that codify relational behaviour between entities in knowledge bases. KGs can simultaneously model many different types of subject-predicate-object and higher-order relations. As such, they offer a flexible modeling framework that has been applied to many areas, including biology and pharmacology -- most recently, in the fight against COVID-19. The flexibility of KG modeling is both a blessing and a challenge from the learning point of view. In this paper we propose a novel coupled tensor-matrix framework for KG embedding. We leverage tensor factorization tools to learn concise representations of entities and relations in knowledge bases and employ these representations to perform drug repurposing for COVID-19. Our proposed framework is principled, elegant, and achieves $100\%$ improvement over the best baseline in the COVID-19 drug repurposing task using a recently developed biological KG.

\end{abstract}
% Keywords command
\providecommand{\keywords}[1]
{
  \small	
  \textbf{\textit{Keywords---}} #1
}
\section{Introduction}
% entities nodes relations edges

How does COVID-19 relate to better-studied viral infections and biological mechanisms? Can we use existing drugs to effectively treat COVID-19 symptoms? Since the COVID-19 pandemic has disrupted our lives, there is a pressing need to answer such  questions, and COVID-19 research has swiftly risen to the top of the scientific agenda, worldwide. While these questions will ultimately be answered by medical experts, data-driven methods can help to cut-down the immense search space, thus helping to accelerate progress and optimize the allocation of precious research resources. In this paper, our goal is to derive such a method by using network science and multi-view analysis tools.

 Networks are powerful abstractions that model interactions between the entities of a system \cite{barabasi2016network}. Networks and network science offer concise and informative modeling, analysis and processing for various biological, engineering and social systems, to name a few \cite{easley2010networks,newman2018networks}. Networks are usually represented by graphs, that are defined by a set of nodes and a set of edges connecting pairs of nodes. The entities of a system are usually represented by the nodes of the graph, and the interactions by the edges.

A knowledge graph (KG) models the relational behavior of various entities in knowledge bases. A KG is heterogeneous in the sense that it models interactions between entities of different type, e.g., drugs and diseases, and is also a multidimensional network (edge-labeled multi-graph) \cite{berlingerio2011foundations}, since the edges (interactions) that connect the nodes (entities) can be multiple and also of different type. Knowledge graphs (KGs) have recently attracted significant attention due to their applicability to various science and engineering tasks. For instance, popular knowledge graphs are YAGO \cite{suchanek2007yago}, DBpedia \cite{auer2007dbpedia}, NELL \cite{carlson2010toward}, Freebase \cite{bollacker2008freebase}, and the
Google KG \cite{singhal2012introducing}. A recent trend codifies knowledge bases of biomedical components and processes, such as genes, diseases and drugs into KGs e.g., \cite{himmelstein2015heterogeneous,himmelstein2017systematic,drkg2020}. KGs can model any relations of the form subject-predicate-object, as well as higher-order generalizations. However, this broad modeling freedom can sometimes be a challenge, as the entities can be very diverse and the dimensions of the KG can turn prohibitively large. 

A common way to exploit KGs for data mining and machine learning applications is via {\em knowledge graph embedding}. KG embedding aims to extract meaningful low dimensional representations of the entities and the relations present in a KG. A plethora of methods have been proposed to perform KG embedding \cite{kolda2005higher,franz2009triplerank,rendle2010pairwise,jiang2012link,riedel2013relation,bordes2013translating,lin2017learning,yang2014embedding,nickel2011three,sun2019rotate,socher2013reasoning,balazevic2019tucker}. The most popular among them adopt a single-layer perceptron or neural network approach e.g., \cite{bordes2013translating,lin2017learning,yang2014embedding,sun2019rotate, socher2013reasoning}. Various tensor factorization models have also been proposed, e.g., \cite{kolda2005higher,franz2009triplerank,rendle2010pairwise,nickel2011three,balazevic2019tucker}. Matrix factorization is also a tool that has been utilized for KG embedding, e.g., \cite{jiang2012link,riedel2013relation}.

In this paper we propose \texttt{TeX-Graph}, a novel coupled tensor-matrix framework to perform KG embedding. The proposed KG coupled tensor-matrix modeling extracts meaningful information from a set of diverse entities with multi-modal interactions in a principled and concise manner. \texttt{TeX-Graph} avoids modeling inefficiencies in previously proposed tensor models, and relative to neural network approaches it offers a principled and effective way to produce unique KG representations. The proposed framework is used for drug repurposing, a pivotal tool in the fight against COVID-19 and other diseases. Learning concise representations for drug compounds, diseases, and the relations between them, our approach allows for link prediction between drug compounds and COVID-19 or other diseases. The impact is critical. First, compound repurposing enables drug design that drastically reduces the design exploration cycle and the failure rate. Second, it markedly reduces drug development cost, as developing new therapeutic drugs is tremendously expensive.

The contributions of our work can be summarized as follows:
\begin{itemize}
	\item {\bf Novel KG modeling}:  We propose a principled coupled tensor-matrix model tailored to KG needs for efficient and parsimonious representations.
	\item {\bf Analysis:} The \texttt{TeX-Graph} embeddings are unique and permutation invariant, a property which is important for consistency and necessary for interpretability. 
	\item {\bf Algorithm:} We design a scalable algorithmic framework with lightweight updates, that can effectively handle very large KGs.
	\item {\bf Application:} The proposed framework is developed to perform drug repurposing, a pivotal task in the fight against COVID-19.
	\item {\bf Performance:} \texttt{TeX-Graph} achieves $100\%$ performance improvement compared to the best available baseline for COVID-19 drug repurposing using a recently developed COVID-19 KG.
\end{itemize}

\noindent{\bf Reproducibility:} The DRKG dataset used in the experiments is publicly available; we will also release our code upon publication of the paper.

\noindent \textbf{Notation:} Our notation is summarized in Table \ref{tab:TableOfNotationForMyResearch}.
\begin{table}[h]
	\caption{Overview of notation.}
	\centering % to have the caption near the table
	\begin{tabular}{r c p{4.4cm} }
		\toprule
% 		$\mathcal{E}$ & $\triangleq$ & Set of edges\\		
% 		$S_\mathcal{G}$ & $\triangleq$ & $N \times N$ adjacency matrix\\	
% 		 $\mathbf{D}$ & $\triangleq$ & $\mathrm{diag}(\mathbf{1}^T\mathbf{A})$ diagonal degree matrix \\
% 		$\mathbf{E}$ & $\triangleq$ & $N\times F$ matrix of embeddings\\
% 		$\mathbf{e}_i$ & $\triangleq$ & $F \times 1$ mbedding vector of node $v_i$ \\
%		% 		$s_\mathcal{G}(\cdot,\cdot)$ & $\triangleq$ & Node -- to -- node similarity  \\
		$x,y,z$ & $\triangleq$ & scalars \\
		$(m,n)$,~$(h,r,t)$& $\triangleq$ &  ordered tuple\\
		$\bm x, \bm y, \bm z$ &$\triangleq$ & vectors \\
		$\bm A, ~\bm B, ~\bm C$ &$\triangleq$ & matrices \\	
		$\underline{\bm X},\underline{\bm Y},\underline{\bm Z}$ &$\triangleq$ & tensors \\
		$\mathcal{S},\mathcal{S}_n^+,\mathcal{S}_n^-$ & $\triangleq$ & sets\\
% 		$\underline{\bm A}$ & $\triangleq$ & tensor \\
$\bm A(:,f)$ &$\triangleq$ & $f$-th column of matrix $\bm A$ \\
		$\bm A(i,:)$ &$\triangleq$ & $i$-th row of matrix $\bm A$ \\
		$\bm{X}^k$ & $\triangleq$ & $k$-th frontal slab of tensor $\underline{\bm{X}}$\\		
		$\bm{A}^T$ & $\triangleq$ & transpose of matrix $\bm{A}$\\		
		$\lVert\bm{A}\rVert_F$ & $\triangleq$ & Frobenius norm of matrix $\bm{A}$ \\
		$\odot$& $\triangleq$ & Khatri-Rao (columnwise Kronecker) product  \\
		$\circ$& $\triangleq$ & outer product \\
		$\ast$& $\triangleq$ & Hadamard product \\	
		diag$(\bm x)$& $\triangleq$ & diagonal matrix of vector $\bm x$\\	
% 		$r(\A)$, $r_{\A}$& $\triangleq$ & rank of $\A$ \\
% 		$k(\A)$, $k_{\A}$& $\triangleq$ & Kruskal rank of $\A$ \\
		$\lfloor x \rfloor$ & $\triangleq$ &  largest integer that is less than or equal to $x$\\
% 		$\bm I$& $\triangleq$ &  Identity matrix\\
% 		$\bm 1$& $\triangleq$ &  vector of ones\\
% 		$|$& $\triangleq$ &  such that\\
% 		$\in$& $\triangleq$ &  belongs to\\
		nnz& $\triangleq$ &  number of non-zeros\\
		\bottomrule
	\end{tabular}
		\label{tab:TableOfNotationForMyResearch}
\end{table}

\section{Preliminaries}\label{prelim}
In order to facilitate the upcoming discussion we now discuss some tensor algebra preliminaries. For more background on tensors the reader is referred to  \cite{sidiropoulos2017tensor,kolda2009tensor}.

\begin{figure}
	\centering
	\includegraphics[width=0.6\linewidth]{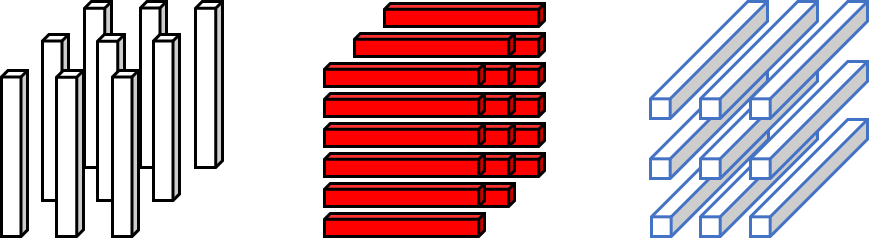}
	\caption{The columns, rows and fibers of a third-order tensor.}
	\label{fig:modes}
\end{figure}

\begin{figure}
	\centering
	\includegraphics[width=0.6\linewidth]{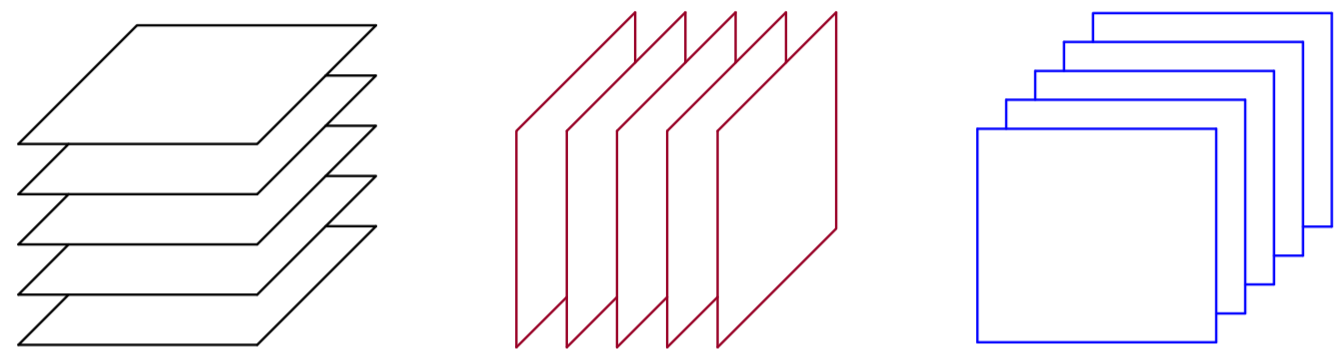}
	\caption{The horizontal, vertical, and frontal slabs of a third-order tensor.}
	\label{fig:modes2}
\end{figure}

A third-order tensor $\underline{\bm X}\in\mathbb{R}^{I\times J\times K}$ is a three-way array indexed by $i,j,k$ with elements $\underline{\bm X}(i,j,k)$. It has three mode types -- columns $\underline{\bm X}(i,:,k)$ ($:$ is used to denote all relevant index values for that mode, i.e., from beginning to end), rows $\underline{\bm X}(:,j,k)$, and fibers $\underline{\bm X}(i,j,:)$ -- see Fig. \ref{fig:modes}. A third order tensor can also be described by a set of matrices (slabs), i.e., horizontal $\underline{\bm X}(i,:,:)$, vertical $\underline{\bm X}(:,j,:)$ and frontal slabs $\underline{\bm X}(:,:,k)$  -- see Fig. \ref{fig:modes2}. A rank-one third order tensor $\underline{\bm Z}\in\mathbb{R}^{I\times J\times K}$ is defined as the outer product of three vectors. Recall that a rank one matrix is the outer product of two vectors.
% A rank-one tensor $\underline{\bm Z}\in\mathbb{R}^{I\times J\times K}$ is the outer product of three vectors, $\bm a\in\mathbb{R}^{I},~\bm b \in\mathbb{R}^{J},~\bm c \in\mathbb{R}^{K}$, denoted as $\underline{\bm Z}=\bm a\circ \bm b\circ \bm c$, where $\circ$ is the outer product operator. 
Any third order tensor can be decomposed into a sum of three-way outer products (rank one tensors) as: 
\begin{equation}\label{PD2}
\underline{\bm X}=\sum_{f=1}^{F} \A(:, f) \circ \B(:, f) \circ \C(:, f),
\end{equation}
where $\bm{A}\in\mathbb{R}^{I\times F},~\bm{B}\in\mathbb{R}^{J\times F},~\bm{C}\in\mathbb{R}^{K\times F}$ are matrices collecting the respective mode factors, i.e., hold in their columns the vectors involved in the $F$ three-way outer products.
The above expression is known as the polyadic decomposition (PD) of a third-order tensor. If $F$ is the minimum number of outer products required to synthesize $\underline{\bm X}$, then $F$ is the \textit{rank} of tensor $\underline{\bm X}$ and the decomposition is known as the \textit{canonical polyadic decomposition} (CPD) or \textit{parallel factor analysis} (PARAFAC) \cite{harshman1994parafac}. For the rest of the paper we use the notation $\underline{\bm X}=\left\llbracket{\bm A},{\bm B},{\bm C}\right\rrbracket$ to denote the CPD of the tensor.  

A striking property of the CPD is that it is essentially unique under mild conditions, even if the rank is higher than the dimensions-- see \cite{chiantini2012generic} for a generic result.

% \begin{Theorem}\label{thm:CPD_generic}
% 	\cite[p.~1019-1021]{chiantini2012generic} 
% 	Let $\underline{\bm X}=\left\llbracket{\bm A},{\bm B},{\bm C}\right\rrbracket$ with $\bm A : I\times F$, $\bm B : J\times F$, and $\bm C : K\times F$. Assume that ${\bm A}$, ${\bm B}$ and ${\bm C}$ are drawn from an absolutely continuous joint distribution {with respect to the Lebesgue measure in $\mathbb{R}^{(I+J+K)F}$}. Also assume $I\geq J\geq K$ without loss of generality. If $F \leq 2^{\lfloor\log_2 J\rfloor+\lfloor\log_2 K\rfloor-2}$, then the decomposition of $\underline{\bm X}$ in terms of $\bm A, \bm B$, and $\bm C$ is essentially unique, almost surely. 
% \end{Theorem}

% Here, essential uniqueness means that if $\tilde{\bm A},\tilde{\bm B},\tilde{\bm C}$ also satisfy $\underline{\bm X}=\llbracket \tilde{\bm A},\tilde{\bm B},\tilde{\bm C}\rrbracket$, then ${\bm A}=\tilde{\bm A}{\bm \Pi}{\bm \Lambda}_1$, ${\bm B}=\tilde{\bm B}{\bm \Pi}{\bm \Lambda}_2$, and ${\bm C}=\tilde{\bm C}{\bm \Pi}{\bm \Lambda}_3$, where ${\bm \Pi}$ is a permutation matrix and ${\bm \Lambda}_i$ is a full rank diagonal matrix such that ${\bm \Lambda}_1{\bm \Lambda}_2{\bm \Lambda}_3={\bm I}$.

A tensor can be represented in a matrix form by employing the \textit{matricization} operation. There are three common ways to matricize (or unfold) a third-order tensor, by stacking columns, rows, or fibers of the tensor to form a matrix. 
To be more precise let:
\begin{equation}
\bm{\underline{X}}(:,:,k)=\bm X^k\in\mathbb{R}^{I\times J},
\end{equation}
where $\bm X^k$ are the frontal slabs of tensor $\bm{\underline{X}}$ and in the context of this paper they model adjacency matrices.
Then the mode-$1$, mode-$2$ and mode-$3$ unfoldings of $\underline{\bm X}$ are:
\begin{equation}{\footnotesize
\bm{{X}^{(1)}}=\begin{bmatrix}
\bm X^1, \hdots, \bm X^K
\end{bmatrix}^T=({\bm C}\odot{\bm B}){\bm A}^T\in\mathbb{R}^{JK\times I},}
\end{equation}
\begin{equation}{\footnotesize
\bm{{X}^{(2)}}=\begin{bmatrix}
\bm X^{1^T}, \hdots, \bm X^{K^T}
\end{bmatrix}=({\bm C}\odot{\bm A}){\bm B}^T\in\mathbb{R}^{IK\times J},}
\end{equation}
\begin{equation}{\footnotesize
\bm{{X}^{(3)}}=\begin{bmatrix}
\bm X^1(:,1), \cdots, \bm X^K(:,1)\\
% \bm X_1(:,2), \cdots, \bm X_K(:,2)\\
\vdots\\ \bm X^1(:,J), \cdots, \bm X^K(:,J)
\end{bmatrix}=({\bm B}\odot{\bm A}){\bm C}^T\in\mathbb{R}^{IJ\times K},}
\end{equation}

Another important tensor model is the coupled CPD. In coupled CPD we are interested in decomposing an array of tensors that share at least one common latent factor. In particular, consider a collection of $N$ tensors $\underline{\bm X}_n\in\mathbb{R}^{I\times J_n\times K_n},~n\in\{1,\dots N\}$. The rank-$F$ coupled CPD of $\{\underline{\bm X}_n\}$ can be expressed as:
\begin{equation}\label{eq:coupledCPD}
    \underline{\bm X}_n=\left\llbracket{\bm A},{\bm B}_n,{\bm C}_n\right\rrbracket,~n\in\{1,\dots N\},
\end{equation}
where $\bm A\in\mathbb{R}^{I\times F}$ is the common factor and $\bm B_n\in\mathbb{R}^{J_n\times F},~\bm C_n\in\mathbb{R}^{K_n\times F}$ are unshared factors. The coupled CPD is also unique under certain conditions, even if individual CPDs of $\bm X_n$ are not unique. In this work we will use the following uniqueness theorem for coupled CPD:

\begin{Theorem}\label{thm:coupled_CPD}
	\cite[p.~510]{sorensen2015coupled} 
	Consider the coupled CPD as expressed in \eqref{eq:coupledCPD}. The decomposition is essentially unique if:
	\begin{enumerate}
	    \item $\bm A$ has full column rank,
	    \item $\bm G$ has full column rank,
	\end{enumerate}
\end{Theorem}
where $\G$ is defined as:
\begin{align}
    \bm G&=\begin{bmatrix}
    C_2\left(\bm C_1\right)\odot C_2\left(\bm B_1\right)\\
    \vdots\\
    C_2\left(\bm C_N\right)\odot C_2\left(\bm B_N\right)
    \end{bmatrix}\\&\in\mathbb{R}^{\frac{1}{4}\sum_{n=1}^NJ_n(J_n-1)K_n(K_n-1)\times \frac{1}{2}F(F-1)},\nonumber
\end{align}
and $C_2\left(\bm C_n\right)$ is the second compound matrix of $\C_n$-- see \cite[p.~861-862]{domanov2013uniqueness}.
In the context of coupled CPD, essential uniqueness corresponds to $\bm A$ being unique and $\{\bm B_n,~\bm C_n\}$ being identifiable up to column scaling and counter-scaling.

%With the above definition, we introduce SRHT.

\section{Problem Statement}

\begin{figure}%{R}{2.2in}
	% 	\vspace{-0.5cm}
	\begin{center}
		\includegraphics[width=2.2in]{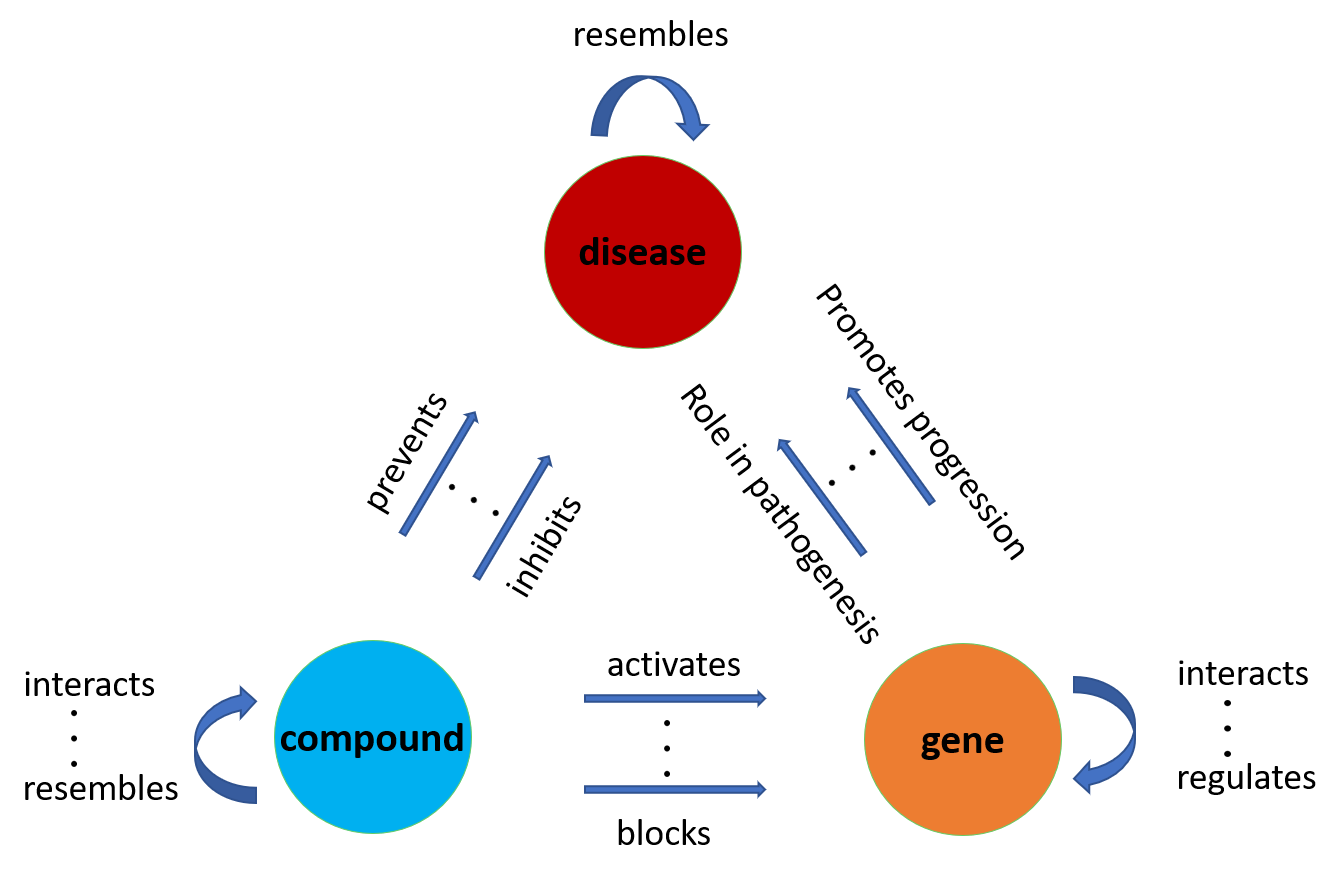}
		\caption{\small Schematic representation of biological KG.}
		\label{fig:KG_modeling}
	\end{center}
	% \vspace{-0.5cm}
\end{figure}

As mentioned in the introduction knowledge graphs (KGs) have attracted significant attention over the past decade due to their tremendous modeling capabilities. In particular, KGs model triplets of subject-predicate-object, denoted as (head, relation, tail) or (h, r, t). Subjects (heads) and objects (tails) are entities that are represented as graph nodes and predicates (relations) define the type of edge according to which the subject is connected to the object. A schematic representation of a KG, which models relations between genes, compounds and diseases is presented in Fig. \ref{fig:KG_modeling}.

In this paper, we focus our attention on a biological KG that models relational triplets between biological entities. For example, (compound 1, interacts with, compound 2), (compound 1, activates, gene 1), (gene 1, regulates, gene 2), (compound 1, prevents, disease 1), (gene 2, is linked with, disease 2) are common triplets in numerous recently developed knowledge bases \cite{himmelstein2015heterogeneous,himmelstein2017systematic,drkg2020}. Modeling these types of relations as a KG enables embedding entities and relations in a Euclidean space which can further facilitate any type of processing and analysis. For instance, obtaining a low dimensional representation of compounds, diseases and the `prevents' relation allows measuring similarity, and thus predicting and testing hypotheses regarding (compound, prevents, disease) interactions. Drug repurposing can be performed by predicting candidate compounds for new and existing target diseases.
Note that the proposed framework to be introduced shortly is not limited to biological KGs -- it can be applied to a wide variety of interesting KGs. 

\subsection{Prior Art}

Several methods have been proposed to learn low dimensional representations of KGs. To properly describe them we need to define the score function $f(\cdot)$ and the loss function $\mathcal{L}(\cdot)$.

Let $(h_n,r_n,t_n)$ be an available triplet and $\bm h_n\in\mathbb{R}^{F},~\bm t_n\in\mathbb{R}^{F}$ and $\bm r_n\in\mathbb{R}^{d}$ be the low dimensional embeddings we aim to learn. Note that entity and relation embeddings need not be of the same dimension. The score function determines the relation model between the head (subject) and the tail (object). In simple words, high values of the score function $f(\bm h_n,\bm r_n,\bm t_n)$ are desirable for existing triplets $(h_n,r_n,t_n)$ and low values of $f(\bm h_n,\bm r_n,\bm t_n)$ for non-existing ones.

In order to produce the entity and relational embeddings we define the following forward model for each triplet $(h_n,r_n,t_n)$:
\begin{subequations}
\begin{equation}
    \bm h_n = \gamma\left(\bm W_e^T \bm o^h_n\right)\in\mathbb{R}^{F},
\end{equation}
\begin{equation}
    \bm t_n = \gamma\left(\bm W_e^T \bm o^t_n\right)\in\mathbb{R}^{F},
\end{equation}
\begin{equation}
    \bm r_n= \delta\left(\bm W_r^T \bm o^r_n\right)\in\mathbb{R}^{d},
\end{equation}
\end{subequations}
where $\bm o^h_n\in\mathbb\{0,1\}^{L_e},~\bm o^t_n\in\mathbb\{0,1\}^{L_e},~\bm o^r_n\in\{0,1\}^{K_r}$ are one-hot input vectors corresponding to the head, tail and relation index of the triplet $(h_n,r_n,t_n)$ respectively, with $L_e,~K_r$ being the total number of entities (nodes) and types of relations; $\gamma(\cdot)$ and $\delta(\cdot)$ are element-wise functions and $\bm W_e\in\mathbb{R}^{L_e\times F},~\bm W_r\in\mathbb{R}^{K_r\times d}$ are matrices that contain the model parameters to be learned.

% Let $\bm A\in\mathbb{R}^{L_e\times F}$ be the matrix with rows the $F-$dimensional embeddings of $L_e$ entities of any type and $\bm C\in\mathbb{R}^{K_r\times d}$ be the matrix with rows the $d-$dimensional embeddings of $K_r$ relations. Then the representation for a triplet $(h_n,r_n,t_n)$ is described by the following set of equations:
% \begin{subequations}
% \begin{equation}
%     \bm h_n =\bm A^T\bm o^h_n = g\left(\bm W_e^T \bm o^h_n\right)\in\mathbb{R}^{F},
% \end{equation}
% \begin{equation}
%     \bm t_n = \bm A^T\bm o^t_n = g\left(\bm W_e^T \bm o^t_n\right)\in\mathbb{R}^{F},
% \end{equation}
% \begin{equation}
%     \bm r_n =\bm C^T\bm o^r_n = h\left(\bm W_r^T \bm o^r_n\right)\in\mathbb{R}^{d},
% \end{equation}
% \end{subequations}
% where $g(\cdot)$ and $h(\cdot)$ are element-wise functions, $\bm W_e\in\mathbb{R}^{L_e\times F},~\bm W_r\in\mathbb{R}^{K_r\times d}$ are the parameter matrices, and $\bm o^h_n\in\mathbb\{0,1\}^{L_e},~\bm o^t_n\in\mathbb\{0,1\}^{L_e},~\bm o^r_n\in\{0,1\}^{K_r}$ are one-hot vectors corresponding to the head, tail and relation of the triplet $(h_n,r_n,t_n)$ respectively. 

 Popular choices for $\gamma(\cdot)$ and $\delta(\cdot)$ are the identity function and hyperbolic tangent. If $\gamma(\cdot)$ or $\delta(\cdot)$ are identity functions then the rows of $\bm W_e$ or $\bm W_e$ are the learned embeddings for entities and relations respectively. For TransE, DistMult and RotatE $F=d$, whereas for TransR and RESCAL $d\neq F$. In the TransR model $\bm M_r\in\mathbb{R}^{d\times F}$ is a projection matrix associated with relation r and in RESCAL $\bm R\in\mathbb{R}^{F\times F}$. 
\begin{table}
	\caption{Knowledge Graph models}
	\label{tab: score functions}
	\begin{tabular}{lc}
	\hline
	Model & score function $f(\bm h, \bm r, \bm t)$ \\ 
	\hline
	TransE \cite{bordes2013translating}& $1-\lVert \bm h+\bm r -\bm t\rVert_2$ or $1-\lVert \bm h+\bm r -\bm t\rVert_1$  \\
	TransR \cite{lin2017learning}& $1-\lVert \bm M_r\bm h+\bm r -\M_r\bm t\rVert_2$  \\
	DistMult\cite{yang2014embedding} &  $\bm h^T$diag$(\bm r)\bm t$\\
		RESCAL \cite{nickel2011three} &  $\bm h^T\bm R\bm t$\\
	RotatE \cite{sun2019rotate}&  $1-\lVert\bm h\ast \bm r-\bm t\rVert$\\
	\hline
	\end{tabular}
\end{table}

In order to learn the embeddings, state-of-the-art methods attempt to minimize the following risk:
\begin{align}\label{empirical}
    \minimize_{\bm W_e,\bm W_r} \frac{1}{N}\sum_{n=1}^N \mathcal{L}\left(y_n-f(\bm h_n, \bm r_n,\bm t_n)\right)
\end{align}
where $N$ is the number of data points (triplets or non-triplets), $y_n=1$ if the triplet $(h_n,r_n,t_n)$ exists, else $y_n=0$. Typical loss functions include the logistic loss, square loss, pairwise ranking loss, margin-based ranking loss and variants of them. In order to tackle the problem in \eqref{empirical} the most popular approach is stochastic gradient descent (SGD) or batch SGD \cite{bottou2010large}.
%\noindent {\large {\bf (T1) Tensor embedding of conventional knowledge graphs}}
\subsection{The 3-way model}\label{subsect:3way}
Modeling a KG using a third order tensor has been considered in  \cite{kolda2005higher,franz2009triplerank,rendle2010pairwise,nickel2011three,balazevic2019tucker}. 
% \cite{kolda2005higher,franz2009triplerank,nickel2011three}. 
In these works, the first and second mode of the tensor is the concatenation of all the available entities, regardless of their type, whereas the third mode represents the different type of relations -- i.e., each frontal slab of the third order tensor represents a certain interaction type between the entities of the KG. The methods in \cite{rendle2010pairwise,balazevic2019tucker} work with incomplete tensors, whereas \cite{kolda2005higher,franz2009triplerank,nickel2011three} model each frontal slab as an adjacency matrix.
% \cite{kolda2005higher,franz2009triplerank,nickel2011three} represent unobserved triplets as non-edges, whereas \cite{rendle2010pairwise,nickel2011three} as unknown.
To be more precise, let $\underline{\bm Z}\in\{0,1\}^{L_e\times L_e\times K_r}$ be the third order tensor in \cite{kolda2005higher,franz2009triplerank,nickel2011three}. Then $\underline{\bm Z}(i,j,k)=1$ if entity $i$ interacts with entity $j$ through relation $k$ and $\underline{\bm Z}(i,j,k)=0$ if there is no interaction between entities $i$ and $j$ via the $k$ relation.

An important observation is that although the first and second mode of tensor $\underline{\bm Z}$ represent the same entities, each frontal slab $\bm Z^k$ is not necessarily symmetric. The reason is that subject-predicate-object does not necessarily imply object-predicate-subject. The works in \cite{kolda2005higher,franz2009triplerank} compute the CPD of $\underline{\bm Z}$ (or scaled versions of $\underline{\bm Z}$) and produce two embeddings for each entity, one as a subject and another as  an object. Although this is not always a drawback, it can result in an overparametrized model because in many applications entities usually act {\em either} as a subject {\em or} as an object, but not both. Furthermore, a single unified representation is usually preferable. In order to overcome this issue, RESCAL \cite{nickel2011three} proposed the following model for each frontal slab:
\begin{equation}\label{rescal}
    \bm Z^k = \bm A \bm R^k \bm A^T,~~~k=1,\dots,K_r,
\end{equation}
where $\bm R^k\in\mathbb{R}^{F\times F}$ is square matrix holding the relation embeddings associated with relation $k$. Note that the RESCAL model is different than the traditional CPD (symmetric in mode $1$ and $2$) in the sense that $\bm R^k$ is not constrained to be diagonal. Relaxing the diagonal constraints allows matrix $\bm R^k$ to absorb in the relation embedding the direction in which different entities interact. 
% However, this type of relaxation in the $\bm R^k$ matrices induces two important drawbacks. First, the number of parameters is increased compared to , thus the KG modeling is prone to over-fitting. Second, it is unclear whether or not the RESCAL model is identifiable, i.e., produces unique embeddings. 
On the downside, this type of relaxation forfeits the parsimony and uniqueness properties of the CPD. This is an important point, since uniqueness is a prerequisite for model interpretability when we are interested in exploratory / explanatory analysis (and not simply in making `black box' predictions).   
% In simple words, although the RESCAL model can handle the asymmetric nature of triplet interactions, it looses the parsimony and uniqueness characteristics of the CPD.

Another important drawback of the tree-way model is that it models unnecessary interactions. To see this, consider a KG that describes interactions between genes and diseases. Suppose that the observed interactions are of gene-gene and gene-disease type but there are no available data for disease-disease interactions. The tree-way model involves disease-disease interactions in the learning process (as non-edges), even though there are no data to justify it. As we will see in the upcoming section our proposed coupled tensor-matrix modeling addresses all the aforementioned challenges.

% model interactions for which we do not have data
% symmetrizes every relationship. While symmetrizing interactions between same entity types seems reasonable. Symmetrizing relationships of different types is not. If we do not symmetrize these relashionships we ran a semi symmetric CPD of an unsymmetric tensor which is not a good idea.

\section{The \texttt{TeX-Graph} model}
\begin{figure}[ht]
	\centering

	\includegraphics[width=0.45\textwidth]{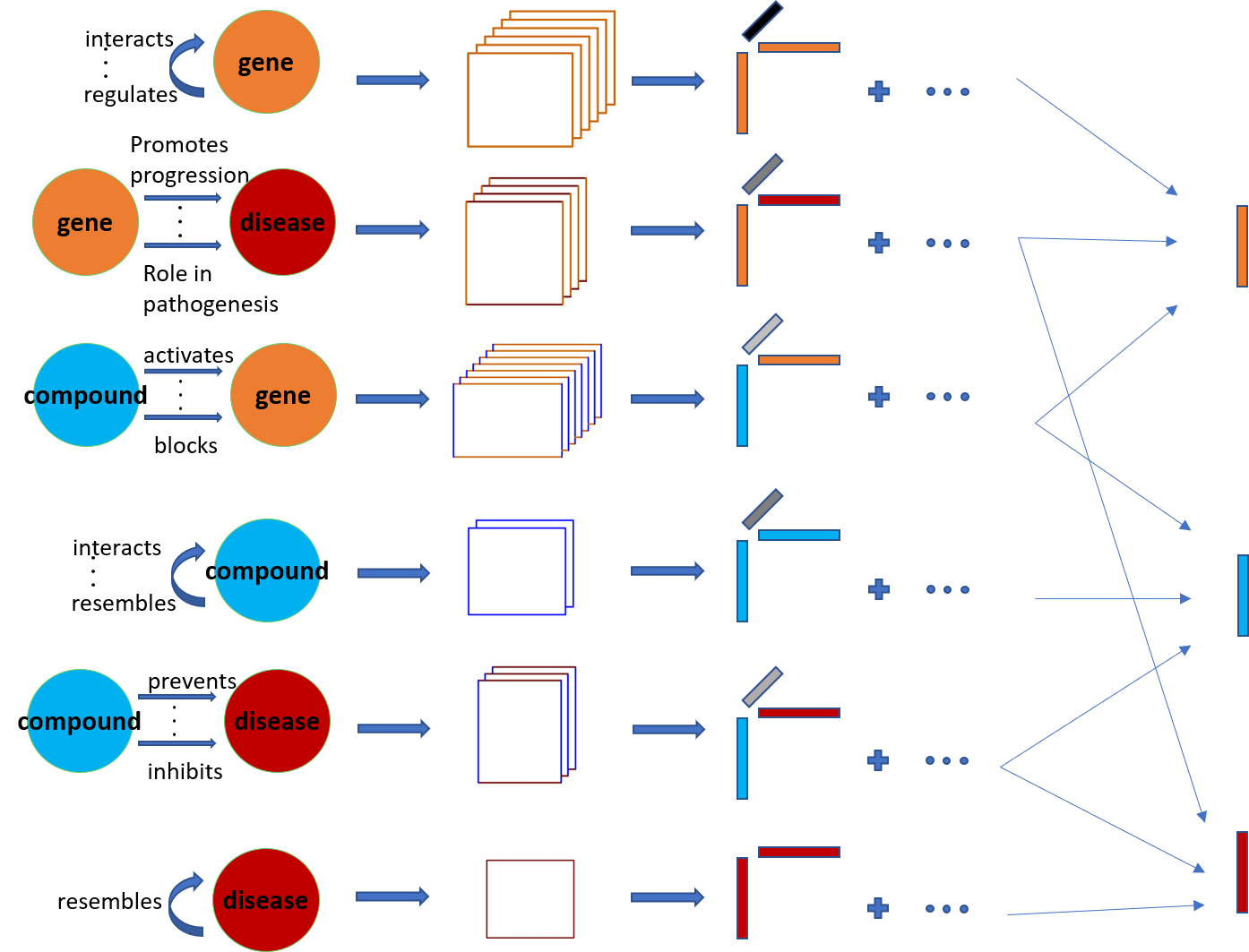}
	\caption{Schematic representation of \texttt{TeX-Graph} model.}
	\label{fig:tensor_modeling}
\end{figure}

In this paper we leverage coupled tensor-matrix factorization to extract low dimensional representations of entities (head, tail) as well as representations for the interactions (relation). KGs can be naturally represented by a collection of tensors and matrices, as shown in Fig. \ref{fig:tensor_modeling}. To see this, consider the previous example of gene, compound and disease entities. Gene-compound interactions, of a certain type, can be represented by an adjacency matrix. Since there are multiple types of interactions, multiple adjacency matrices are necessary to model every interaction, resulting in a tensor $\bm{\underline{\bm X}}_{g,c}\in{\{0,1\}}^{L_g\times L_c\times K_{g,c}}$, where $L_g,~L_c$ are the number of genes and compounds respectively, and $K_{g,c}$ is the number of different interactions between genes and compounds. The same idea can be applied to any 
% combination of available gene, compound and disease interactions in our example and any
(entity,interaction,entity) triplet.

To facilitate the discussion let $\underline{\bm X}_{m,n}\in{\{0,1\}}^{L_m\times L_n\times K_{m,n}}$ be the tensor of interactions between entity of type-$m$ and type-$n$, e.g., $m$ codifies genes and $n$ codifies compounds. Also let $L_{T}$ be the total number of different entity types, then $m,n\in\{1,\dots,L_{T}\}$. $\underline{\bm X}_{m,n}(i,j,k)=1$ if the $i$-th entity of type-$m$ interacts with the $j$-th entity of type-$n$ via relation $k$ and $\underline{\bm X}_{m,n}(i,j,k)=0$ if there is no type-$k$ interaction between the $i$-th entity of type-$m$ and the $j$-th entity of type-$n$. The KG is represented by a collection of tensors as: 

\begin{align}
&\underline{\bm X}_{m,n}\in{\{0,1\}}^{L_m\times L_n\times K_{m,n}},~(m,n)\in\mathcal{S}\\
&\mathcal{S}=\{(m,n): ~m\leq n,~\nonumber\\&\exists~(h,r,t)~\text{with}~(h,t)~\in\text{type}~(m,n) ~\text{or} ~(n,m)\}\nonumber,
\end{align}
where $\sum_{n=1}^{L_T}L_n=L_e$ and $\sum_{(m,n)\in\mathcal{S}}K_{m,n}=K_r$.
Note that tensors $\underline{\bm X}_{m,n}$ and $\underline{\bm X}_{n,m}$ contain the same information since ${\bm X}_{m,n}^k={\bm X}_{n,m}^{k^T}$. Therefore we only consider $(m,n)$ tuples where $m\leq n$.

Each of the tensors in the array $\{\underline{\bm X}_{m,n},~(m,n)\in\mathcal{S}\}$ admits a CPD and the overall model is cast as:
\begin{equation}\label{model:coupledKG}
    \underline{\bm X}_{m,n}=\left\llbracket{\bm A_m},{\bm A_n},{\bm C_{m,n}}\right\rrbracket,~(m,n)\in\mathcal{S},
\end{equation}
where $\bm A_n\in\mathbb{R}^{L_n\times F},\bm C_{m,n}\in\mathbb{R}^{K_{m,n}\times F}$. The $i$-th row of $\bm A_n$ represents the $F$-dimensional embedding of the $i$-th type-$n$ entity and the $k$-th row of $\bm C_{m,n}$ represents the $F$-dimensional embedding of the $k$-th type relation between type-$m$ and type-$n$ entities. Note that in the case where entities of type-$m$ interact with entities of type-$n$ via only one type of relation, $\bm X_{m,n}\in{\{0,1\}}^{L_m\times L_n}$ is a matrix and can be factored as:
\begin{equation}
    {\bm X}_{m,n}={\bm A_m}\text{diag}(\bm c_{m,n}){\bm A_n}^T
\end{equation}

The model in \eqref{model:coupledKG} is a coupled CPD as the factors $\bm A_n$ appear in multiple tensors. For instance, type-1-type-1 interactions (gene-gene), type-1-type-2 interactions (gene-compound), type-1-type-3 interactions (gene-disease), result in the factor $\bm A_1$ appearing in tensors $\underline{\bm X}_{1,1}=\left\llbracket\bm A_1,\bm A_1,\bm C_{1,1}\right\rrbracket,\underline{\bm X}_{1,2}=\left\llbracket\bm A_1,\bm A_2,\bm C_{1,1}\right\rrbracket$ and $\underline{\bm X}_{1,3}=\left\llbracket\bm A_1,\bm A_3,\bm C_{1,3}\right\rrbracket$.

The proposed \texttt{TeX-Graph} exhibits several favorable properties. First, the produced embeddings are unique, provided that they appear in more than one adjacency matrices.

\begin{Prop}{\emph{(Uniqueness of the embeddings)}}\label{proposition}
	If the coupled tensor model in \eqref{model:coupledKG} is indeed low-rank, $F$, there exist entity and relation embedding vectors in $F$ dimensional space that generate the given knowledge base. Then the $F-$dimensional $\texttt{TeX-Graph}$ embeddings for type-$n$ entities and type-$(m,n)$ relations are unique and permutation invariant provided that $\sum_{m\in \mathcal{S}_n^+} K_{m,n}+\sum_{p\in \mathcal{S}_n^-}K_{n,p}>1$ and $K_{m,n}>1$ respectively, where $\mathcal{S}_n^+,~\mathcal{S}_n^-$ are defined in \eqref{eq:subsets}.
\end{Prop}

The proof of Proposition \ref{proposition} utilizes the uniqueness results of Theorem \ref{thm:coupled_CPD} and is relegated to the journal version due to space limitations. In the case where $K_{m,n}=1$ and type-$m$ entities appear in multiple tensors but type-$n$ entities only in one, the \texttt{TeX-Graph} model identifies $\bm A_m$ and $\bm A_n$diag$(\bm c_{m,n})$, since there is rotational freedom between $\bm A_n$ and $\bm c_{m,n}$.

Another important property of the proposed \texttt{TeX-Graph} is that it avoids modeling of spurious `cross-product' relations that can never be observed. The coupled tensor-matrix model allows for a concise KG representation that eliminates such spurious relations from the start, contrary to the three-way model. To see this, consider the previous example of gene-disease KG that observes relational triplets between gene-gene and gene-disease type but not for disease-disease type. The proposed \texttt{TeX-Graph} does not model  disease-disease interactions, whereas the three-way model treats them as non-edges. 

It is worth noticing that \texttt{TeX-Graph} makes the implicit assumption that $\underline{\bm X}_{n,n}$ are symmetric in the first and second mode. This is not always the case, since interactions between some entity types might be directed. To overcome this issue we assume that (h,r,t) implies (t,r,h) for (h,t) of the same type. Although this assumption ignores the direction in this type of interactions, it results in a more parsimonious model for the entity embeddings.   

\subsection{Algorithmic framework}\label{sec:Algorithms}
In order to learn the $F$-dimensional embeddings of all entities and relations we formulate the KG embedding problem as: 
\begin{align}\label{coupledKG}
    \minimize_{\substack{ \{{\bm A}_m\},   \{{\bm C}_{m,n}\} }} ~\sum_{(m,n)\in\mathcal{S}}\left\|{\underline{\bm X}}_{m,n}-\llbracket{\bm A_m},{\bm A_n},{\bm C_{m,n}}\rrbracket\right\|_F^2,
\end{align}
The problem in \eqref{coupledKG} is non-convex and NP-hard in general. In order to tackle it we propose to fix all variables but one and update the remaining variable. This procedure is repeated in an alternating fashion. The update for $\bm A_n$ is a system of linear equations and takes the form:
% {\footnotesize\begin{align}\label{eq:Aupdate}
%     \sum_{m\in \mathcal{S}_n^+}\left(\bm C_{m,n}\odot\bm A_m\right)^T\bm X_{m,n}^{(2)}+\sum_{p\in \mathcal{S}_n^-}\left(\bm C_{n,p}\odot\bm A_p\right)^T\bm X_{n,p}^{(1)}=\nonumber\\\left(\sum_{m\in \mathcal{S}_n^+}\left(\bm C_{m,n}^T\bm C_{m,n}\ast\bm A_m^T\bm A_m\right)+\sum_{p\in \mathcal{S}_n^-}\left(\bm C_{n,p}^T\bm C_{n,p}\ast\bm A_p^T\bm A_p\right)\right)\bm A_n^T,
% \end{align}}
{\footnotesize\begin{equation}\label{eq:Aupdate}
    \sum_{m\in \mathcal{S}_n^+}\left(\bm C_{m,n}\odot\bm A_m\right)^T\bm X_{m,n}^{(2)}+\sum_{p\in \mathcal{S}_n^-}\left(\bm C_{n,p}\odot\bm A_p\right)^T\bm X_{n,p}^{(1)}=
\end{equation}}
{\scriptsize\begin{equation*}
    \left(\sum_{m\in \mathcal{S}_n^+}\left(\bm C_{m,n}^T\bm C_{m,n}\ast\bm A_m^T\bm A_m\right)+\sum_{p\in \mathcal{S}_n^-}\left(\bm C_{n,p}^T\bm C_{n,p}\ast\bm A_p^T\bm A_p\right)\right)\bm A_n^T,\nonumber
\end{equation*}}
where 
\begin{equation}\label{eq:subsets}
    \mathcal{S}_n^+=\{m:(m,n)\in\mathcal{S}\},~\mathcal{S}_n^-=\{p:(n,p)\in\mathcal{S}\}
\end{equation}
% \[\mathcal{S}_n^+=\{m:(m,n)\in\mathcal{S}\},~\mathcal{S}_n^-=\{p:(n,p)\in\mathcal{S}\}\]
% \textcolor{red}{symmetry}
The update for $\bm C_{m,n}$ is the solution to the following system of linear equations:
\begin{equation}\label{eq:Cupdate}
    \left(\bm A_n\odot\bm A_m\right)^T\bm X_{m,n}^{(3)}=\left(\bm A_{n}^T\bm A_{n}\ast\bm A_m^T\bm A_m\right)\bm C_{m,n}^T
\end{equation}
The derivations for these updates as well as implementation details are presented in Appendix \ref{app:tekgraph}.

The proposed \texttt{TeX-Graph} is presented in Algorithm \ref{algo:TeKGraph}. \texttt{TeX-Graph} is an iterative algorithm that tackles a non-convex problem and NP-hard in general. As a result different initial points might produce different results. Although we have observed that random initialization is sufficient most of the times we propose an alternative initialization procedure that yields consistent and reproducible results. To be more specific we form a symmetric version of tensor $\underline{\bm Z}$ as:
\begin{equation}\label{eq: symmRescal}
    \underline{\bm Y}(i,j,k)=\min\{1,\underline{\bm Z}(i,j,k)+\underline{\bm Z}(j,i,k)\}
\end{equation}
% \[\underline{\bm Y}(i,j,k)=\min\{1,\underline{\bm Z}(i,j,k)+\underline{\bm Z}(j,i,k)\}\]
Then we compute the semi-symmetric CPD of $\underline{\bm Y}=\left\llbracket{\bm A},{\bm A},{\bm C}\right\rrbracket$ using sparse eigenvalue decomposition (EVD) \cite{sanchez1990tensorial}. 
% The details of the algorithm can be found in the supplementary material. 
The proposed initialization procedure is presented in Algorithm \ref{algo:initialization}.
\begin{algorithm}[tb]
	\caption{\texttt{TeX-Graph}}
	\label{algo:TeKGraph}
	\begin{algorithmic}
		\STATE {\bfseries Input:} $\{(h_n,r_n,t_n)\}_{n=1}^N,~\{{\bm A}_m\},~\{{\bm C}_{m,n}\}$.
		\STATE {\bfseries Output:} $\{{\bm A}_n\}_{n=1}^{L_e},~\{{\bm C}_{m,n}\}_{(m,n)\in\mathcal{S}}$.\\
		\STATE Create $\{\underline{\bm X}_{m,n}\}_{(m,n)\in\mathcal{S}}$ from $\{(h_n,r_n,t_n)\}_{n=1}^N$
		\REPEAT
		\FOR{$n\in\{1,\dots, L_E\}$}
		 \STATE\quad$\bm A_n\leftarrow~$ \text{solve}~ \eqref{eq:Aupdate}
		\ENDFOR
		\FOR{$(m,n)\in\mathcal{S}$}
		 \STATE\quad $\bm C_{(m,n)}\leftarrow~$ \text{solve}~ \eqref{eq:Cupdate}
		\ENDFOR
		\UNTIL criterion is met.\end{algorithmic}
\end{algorithm}

\begin{algorithm}[tb]
	\caption{\texttt{TeX-Graph-initialization}}
	\label{algo:initialization}
	\begin{algorithmic}
		\STATE {\bfseries Input:} $\{(h_n,r_n,t_n)\}_{n=1}^N$.
		\STATE {\bfseries Output:} $\{{\bm A}_n\}_{n=1}^{L_e},~\{{\bm C}_{m,n}\}_{(m,n)\in\mathcal{S}}$.\\
		\STATE Create tensor $\underline{\bm Z}$ from $\{(h_n,r_n,t_n)\}_{n=1}^N$ as explained in section \ref{subsect:3way};
		\STATE Form $\underline{\bm Y}$ as: $\underline{\bm Y}(i,j,k)=\min\{1,\underline{\bm Z}(i,j,k)+\underline{\bm Z}(j,i,k)\};$
		\STATE Solve $\underline{\bm Y}=\left\llbracket{\bm A},{\bm A},{\bm C}\right\rrbracket$ via sparse EVD;
		\STATE Form $\{{\bm A}_n\}_{n=1}^{L_e},~\{{\bm C}_{m,n}\}_{(m,n)\in\mathcal{S}}$ from $\bm A,~\bm C$.
		\end{algorithmic}
\end{algorithm}
\subsection{Computational complexity analysis}
In terms of memory requirements and computational complexity, the main bottleneck of \texttt{TeX-Graph} lies in instantiating and computing the matricized tensor times Khatri-Rao product (MTTKRP) in the left hand side (LHS) of \eqref{eq:Aupdate} and \eqref{eq:Cupdate}. The number of flops needed to compute the LHS of \eqref{eq:Aupdate} and \eqref{eq:Cupdate} is $\mathcal{O}\left(F\cdot\text{nnz}\left(\sum_{m\in \mathcal{S}_n^+}\underline{\bm X}_{m,n}+\sum_{p\in \mathcal{S}_n^-}\underline{\bm X}_{n,p}\right)\right)$ and $\mathcal{O}\left(F\cdot\text{nnz}\left(\underline{\bm X}_{m,n}\right)\right)$ respectively. For small values of $F$ which is usually the case in practice the complexity is linear in the number of triplets participating in each update. Furthermore the Khatri-Rao products in the (LHS) of \eqref{eq:Aupdate} and \eqref{eq:Cupdate} are not being instantiated as shown in Appendix \ref{app:tekgraph}.

\section {Drug Repurposing for COVID-19}\label{sec:num}
In this section we apply \texttt{TeKGraph} to a recently developed KG \cite{drkg2020} in order to perform drug repurposing for COVID-19 disease. All algorithms were implemented in Matlab or Python, and executed on a Linux server
comprising 32 cores at 2GHz and 128GB RAM.
\subsection{Data}
The dataset used in the experiments is the Drug Repurposing Knowledge Graph (DRKG)\footnote{github.com/gnn4dr/DRKG} \cite{drkg2020}. It codifies triplets of biological interactions between 97,238 different entities of 13 types, namely, genes, compounds, diseases, anatomy, tax, biological process, cellular component, pathway, molecular function, anatomical therapeutic chemical (Atc), side effect, pharmacological class, and symptom. The total number of triplets is 5,874,258 and there are 107 different types of interactions. The KG is organised in 6 adjacency tensors and 11 adjacency matrices. Detailed description of the dataset and the modeling can be found in Table \ref{tab:freq}. Each row denotes a different adjacency tensor or matrix and $\#$ type-$m$ entities, $\#$ type-$m$ entities, $\#$ relation types correspond to the dimension of mode 1, mode 2, and mode 3 respectively. The last column (sparsity) denotes the sparsity of each tensor, i.e., $\frac{\text{nnz}\left(\underline{\bm X}_{m,n}\right)}{L_m L_n K_{m,n}}$

	\begin{table*}
		\caption{Coupled tensor-matrix DRKG modeling.}
		\label{tab:freq}
		\scalebox{0.77}{
		\begin{tabular}{ccccccc}
			\toprule
			entity type-m&entity type-n& \# type-m entities 1& \# type-n entities 2 & \# relation types & tensor&sparsity \\
			\midrule
			\multirow{9}{*}{ {Gene}}
			&Gene&39,220&39,220&32&$\underline{\bm X}_{1,1}=\left\llbracket{\bm A}_1,{\bm A}_1,{\bm C}_{1,1}\right\rrbracket$ &$6.12~10^{-5}$\\
			&Compound&39,220&24,313&34&$\underline{\bm X}_{1,2}=\left\llbracket{\bm A}_1,{\bm A}_2,{\bm C}_{1,2}\right\rrbracket$ &$6.50~10^{-6}$ \\
			&Disease&39,220&5,103&15&$\underline{\bm X}_{1,3}=\left\llbracket{\bm A}_1,{\bm A}_3,{\bm C}_{1,3}\right\rrbracket$&$4.13~10^{-5}$  \\
			&{Anatomy}&39,220&400&3&$\underline{\bm X}_{1,4}=\left\llbracket{\bm A}_1,{\bm A}_4,{\bm C}_{1,4}\right\rrbracket$&0.0154  \\
			&Tax&39,220&215&1&${\bm X}_{1,5}={\bm A}_1\text{diag}({\bm c}_{1,5}){\bm A}_{5}^T$ &0.0017 \\
			&Biological Process&39,220&11,381&1&${\bm X}_{1,6}={\bm A}_1\text{diag}({\bm c}_{1,6}){\bm A}_{6}^T$ &0.0013 \\
			&Cellular Component&39,220&1,391&1 &${\bm X}_{1,7}={\bm A}_1\text{diag}({\bm c}_{1,7}){\bm A}_{7}^T$ &0.0013\\
			&Pathway&39,220&1,822&1&${\bm X}_{1,8}={\bm A}_1\text{diag}({\bm c}_{1,8}){\bm A}_{8}^T$ &0.0012 \\
			&Molecular Function&39,220&2,884&1&${\bm X}_{1,9}={\bm A}_1\text{diag}({\bm c}_{1,9}){\bm A}_{9}^T$ &$8.6 10^{-4}$ \\
			\hline
			\multirow{4}{*}{ {Compound}}
			&Compound&24,313&24,313&2&$\underline{\bm X}_{2,2}=\left\llbracket{\bm A}_2,{\bm A}_2,{\bm C}_{2,2}\right\rrbracket$&0.0023 \\
			&Disease&24,313&5,103&10 &$\underline{\bm X}_{2,3}=\left\llbracket{\bm A}_2,{\bm A}_3,{\bm C}_{2,3}\right\rrbracket$&$6.76~10^{-5}$ \\
			&Atc&24,313&4,048&1 &${\bm X}_{2,10}={\bm A}_2\text{diag}({\bm c}_{2,10}){\bm A}_{10}^T$&$1.6~10^{-4}$ \\
			&Side Effect&24,313&5,701&1&${\bm X}_{2,11}={\bm A}_2\text{diag}({\bm c}_{2,11}){\bm A}_{11}^T$&0.0010  \\
			&Pharmacological Class&24,313&345&1&${\bm X}_{2,12}={\bm A}_2\text{diag}({\bm c}_{2,12}){\bm A}_{12}^T$&$1.22~10^{-4}$\\
			\hline
			\multirow{3}{*}{ {Disease}}
			&Disease&5,103&5,103&1 &${\bm X}_{3,3}={\bm A}_3\text{diag}({\bm c}_{3,3}){\bm A}_{3}^T$ &$4.17~10^{-5}$\\
			&Anatomy&5,103&400&1 &${\bm X}_{3,4}={\bm A}_3\text{diag}({\bm c}_{3,4}){\bm A}_{4}^T$&0.0018 \\
			&Symptom&5,103&415&1&${\bm X}_{3,13}={\bm A}_3\text{diag}({\bm c}_{3,13}){\bm A}_{13}^T$ & 0.0016  \\
			\hline
			% {Anatomy}&Gene&400&39,220&3&$\underline{\bm X}_{18}=\left\llbracket{\bm U}_A,{\bm U}_G,{\bm C}_{18}\right\rrbracket$&0.0154  \\
			% \hline
			\bottomrule
			\end{tabular}}
			\end{table*}

\subsection{Procedure}
Drug repurposing refers to the task of discovering existing drugs that can effectively manage certain diseases-- COVID-19 in our study. DRKG codifies relational triplets of (compound,treats,disease) and (compound,inhibits,disease). Therefore drug repurposing in the context of DRKG boils down to predicting new `treats' and `inhibits' edges (links) between compounds and diseases of interest.

We follow the evaluation procedure proposed in \cite{drkg2020}. In the training phase we learn low dimensional representations for the entities and relations, using all the edges in DRKG. In the testing phase, we assign a score to (compound,treats,disease) and (compound,inhibits,disease) triplets according to the scoring function used for training. For the proposed \texttt{TeX-Graph}, the scores assigned to the triplet (hyper-edge) (compound $i$,treats,disease $j$) and (compound $i$,inhibits,disease $j$) are:
\[\text{score}_{i,j,2}=\bm A_2(i,:)\text{diag}\left(\bm C_{2,3}\left(2,:\right)\right)\bm A_2(j,:)^T,\]
\[\text{score}_{i,j,9}=\bm A_2(i,:)\text{diag}\left(\bm C_{2,3}\left(9,:\right)\right)\bm A_2(j,:)^T,\]
since `treats' and `inhibits' relations correspond to the second and ninth frontal slab of $\underline{\bm X}_{2,3}$, respectively. The testing set consists of 34 corona-virus related diseases, including SARS, MERS and SARS-COV2 and $8,103$ FDA-approved drugs in Drugbank. Drugs with molecule weight less than 250 daltons are excluded from testing. Ribavirin was also excluded from the testing set, since there exist a `treat' edge in the training set between Ribavirin and a target disease. 
% Thus we can assume that Ribavirin is candidate drug for COVID-19. 
In order to evaluate the
performance of the proposed \texttt{TeX-Graph} and the alternatives we retrieve the top-100 ranked drugs that appear in the highest testing scoring (hyper-)edges. These are the proposed candidate drugs for COVID-19. Then we assess how many of the 32 clinical trial drugs \footnote{www.covid19-trials.com} (Ribavirin is excluded) appear in the proposed candidate top-100 drugs.
\subsection{Methods}
The methods used in the experiments are:
			
			\begin{itemize}
			\item \textbf{TeX-Graph}. The proposed \texttt{TeKGraph} algorithm initialized with Algorithm \ref{algo:initialization}. The embedding dimension is set to $F=50$ and the algorithm runs for 10 iterations.
				\item \textbf{TransE-DRKG} \cite{bordes2013translating,drkg2020}. \texttt{TransE} learns low dimensional KG embeddings using the score function shown in Table \ref{tab: score functions}. For the the task of drug repurposing we use the specifications proposed in \cite{drkg2020}. The $l_2$ norm is chosen in the score function and training is performed using the deep graph library for knowledge graphs \cite{zheng2020dgl}. To evaluate the performance of TransE-DRKG on the drug repurposing task we used the $400-$dimensional pretrained embeddings in \cite{drkg2020}, with which the drug repurposing results were better than the stand-alone code without pretraining.
				 \item \textbf{3-way KG embeddings (3-way KGE)}. We add as a baseline the embeddings produced by computing the CPD of tensor $\underline{\bm Y}$ in \eqref{eq: symmRescal}. Recall that we use an algebraic CPD of $\underline{\bm Y}$ to initialize \texttt{TeX-Graph}. In \texttt{3-way KGE} we initialize using the same procedure and also run 10 alternating least-squares iterations to compute the CPD of $\underline{\bm Y}$. \texttt{3-way KGE} is tested with $F=50$.
			\end{itemize}
% 			We also tested RESCAL in our experiments. The SGD version failed to produce satisfactory results and the ALS version was exhausting all the resources, therefore omitted from the paper presentation.
			% \subsection{Node classification}
\subsection{Results}		
Table \ref{tab:covidgrugs} shows the clinical trial drugs that appear in the top-100 recommendations along with their [rank-order]. The proposed approach retrieves 10 clinical trial drugs in the top-100 positions, and 7 in the top-50. Compared to \texttt{TransE-DRKG} that was the first proposed algorithm to perform drug-repurposing for COVID-19, \texttt{TeX-Graph} achieves $75\%$ and $100\%$ improvement in precision in the top-$50$ and top-$100$ respectively.

 It is worth emphasizing that the proposed \texttt{Tex-Graph} retrieves approximately $1/3$ of the COVID-19 clinical trial drugs, in the top-100, among a testing set of $8,103$ drugs. This result is pretty remarkable and can essentially help cutting down the immense search space of medical research. For instance, consider the case of Dexamethasone, which is retrieved by \texttt{Tex-Graph} in the top ranked position (it admitted the highest score among all $8,103$ drugs). At the onset of the pandemic, the initial guidance for Dexamethasone and other corticosteroids was indecisive. Guidelines from different sources issued either a weak recommendation to use Dexamethasone (with an asterisk that further evidence was required) or a weak recommendation against corticosteroids and Dexamethasone \cite{prescott2020corticosteroids}. However, recent results indicate that treatment with Dexamethasone reduces mortality in patients with COVID-19 \cite{horby2020dexamethasone}. The results of \texttt{Tex-Graph} coalign with the latest evidence and rank Dexamethasone as the top recommended drug. This suggests that our proposed data-driven approach could have essentially contributed in overturning the initial hesitancy to administrate Dexamethasone as a first line treatment.
			
\begin{table}
		\caption{Proposed candidate drugs for COVID-19}
		\label{tab:covidgrugs}
		\scalebox{0.7}{
		\begin{tabular}{ccc}
			\toprule
			\texttt{TeX-Graph}& \texttt{TransE-DRKG}&\texttt{3-way KGE} \\
			\midrule
			F=50&F=400&F=50\\
			\hline
			Dexamethasone [1] &Dexamethasone [4]&Oseltamivir [89]  \\
			Methylprednisolone [6]  &Colchine [8]& \\
			Azithromycin [13] &Methylprednisolone [16]& \\
			Thalidomide [18]&Oseltamivir [49]&\\
			Losartan [41] &Deferoxamine [87] &
			 \\
			 Hydroxychloroquine [47]&&\\
			 Colchine [48]&&\\
			 Oseltamivir[60]&&\\
			 Chloroquine[68]&&\\
			 Deferoxamine [88] &&\\
			\hline
			% {Anatomy}&Gene&400&39,220&3&$\underline{\bm X}_{18}=\left\llbracket{\bm U}_A,{\bm U}_G,{\bm C}_{18}\right\rrbracket$&0.0154  \\
			% \hline
			\bottomrule
			\end{tabular}}
			\end{table}		

\section{Conclusion}
In this paper we proposed a novel coupled tensor-matrix framework for knowledge graph embedding. The proposed model is principled and enjoys several favorable properties, including parsimony and uniqueness. The developed algorithmic framework admits lightweight updates and can handle very large graphs. Finally the proposed \texttt{TeX-Graph} showed very promising results in a timely application to drug repurposing, a task of paramount importance in the fight against COVID-19.
\section{Acknowledgements}
The authors would like to acknowledge Ioanna Papadatou, M.D., Ph.D, for contributing in the medical assessment of the produced results.
\appendix
\section{Appendix: \texttt{TeX-Graph} updates}\label{app:tekgraph}
\texttt{TeX-Graph} solves the following problem
\begin{align}\label{coupledKG2}
    \minimize_{\substack{ \{{\bm A}_m\},   \{{\bm C}_{m,n}\} }} ~\sum_{(m,n)\in\mathcal{S}}\left\|{\underline{\bm X}}_{m,n}-\llbracket{\bm A_m},{\bm A_n},{\bm C_{m,n}}\rrbracket\right\|_F^2.
\end{align}
Then the update for $\bm A_n$ is the solution of:
\begin{align}\label{coupledKG3}
    \minimize_{\substack{ \bm A_n }} ~&\sum_{m\in \mathcal{S}_n^+}\left\|{\underline{\bm X}}_{m,n}-\llbracket{\bm A_m},{\bm A_n},{\bm C_{m,n}}\rrbracket\right\|_F^2+\nonumber\\&\sum_{p\in \mathcal{S}_n^-}\left\|{\underline{\bm X}}_{n,p}-\llbracket{\bm A_n},{\bm A_p},{\bm C_{n,p}}\rrbracket\right\|_F^2,
\end{align}
where $\mathcal{S}_n^+,\mathcal{S}_n^-$ are defined in \eqref{eq:subsets}. Problem \eqref{coupledKG3} can be written as:
\begin{align}\label{coupledKG4}
    \minimize_{\substack{ \bm A_n }} ~&\sum_{m\in \mathcal{S}_n^+}\left\|\bm{{X}}_{m,n}^{(1)}-\left({\bm C_{m,n}}\odot{\bm A_m}\right){\bm A_n}^T\right\|_F^2+\nonumber\\&\sum_{p\in \mathcal{S}_n^-}\left\|\bm{{X}}_{n,p}^{(2)}-\left({\bm C_{n,p}}\odot{\bm A_p}\right){\bm A_n}^T\right\|_F^2.
\end{align}
Taking the gradient of \eqref{coupledKG4} with respect to $\bm A_n$ and setting it to zero yields the equation in \eqref{eq:Aupdate}. The main bottleneck of \eqref{eq:Aupdate} in terms of memory requirements and computational complexity is instantiating the Khatri-Rao products $\left(\bm C_{n,p}\odot\bm A_p\right),\left(\bm C_{m,n}\odot\bm A_m\right)$ and computing the MTTKRP $\left(\bm C_{n,p}\odot\bm A_p\right)^T\bm X_{n,p}^{(1)},~\left(\bm C_{m,n}\odot\bm A_m\right)^T\bm X_{m,n}^{(2)}$. We focus on the computation of:
\begin{equation}\label{eq:MTTKRP}
    \left(\bm C_{n,p}\odot\bm A_p\right)^T\bm X_{n,p}^{(1)}.
\end{equation}
Equation \eqref{eq:MTTKRP} can be equivalently written as:
\begin{align}\label{eq:MTTKRP2}
&\begin{bmatrix}
 \bm A_p \text{diag}\left(\bm C_{n,p}(1,:)\right)\\\vdots\\
 \bm A_p \text{diag}\left(\bm C_{n,p}(K_{n,p},:)\right)\\
\end{bmatrix}^T\begin{bmatrix}
\bm X^{1^T}_{n,p}\\\vdots\\ \bm X^{K_{n,p}^T}_{n,p}
\end{bmatrix}
=\nonumber\\&\sum_{k=1}^{K_{n,p}}\text{diag}\left(\bm C_{n,p}(k,:)\right)\bm A_p\bm X_{n,p}^{k^T}.
\end{align}
It is clear from equation \eqref{eq:MTTKRP2} that $\left(\bm C_{n,p}\odot\bm A_p\right)$ need not be instantiated. Furthermore, the number of flops to compute \eqref{eq:MTTKRP2} is $\mathcal{O}(F\cdot\text{nnz}(\underline{\bm X}_{n,p}))$.
Note that computing $\left(\bm C_{m,n}\odot\bm A_m\right)^T\bm X_{m,n}^{(2)}$ is only different in the fact that the frontal slabs are not transposed, and is thus omitted.

The update for ${\bm C_{m,n}}$ is the solution of: 
\begin{equation}\label{eq:Cupd}
    \minimize_{\substack{ \bm C}_{m,n}} ~\left\|{\underline{\bm X}}_{m,n}-\llbracket{\bm A_m},{\bm A_n},{\bm C_{m,n}}\rrbracket\right\|_F^2,
\end{equation}
or equivalently:
\begin{equation}\label{eq:Cupd2}
    \minimize_{\substack{ \bm C}_{m,n}} ~\left\|{{\bm X}}_{m,n}^{(3)}-\left({\bm A_m}\odot{\bm A_n}\right){\bm C_{m,n}}^T\right\|_F^2.
\end{equation}
Taking the gradient of \eqref{eq:Cupd2} with respect to ${\bm C_{m,n}}$ and setting it to zero yields the equation in \eqref{eq:Cupdate}. The main memory and computation bottleneck of equation \eqref{eq:Cupdate} is computing the MTTKRP. The formula in \eqref{eq:MTTKRP2} can be utilized if $\bm C_{n,p}$ is replaced by $\bm A_n$, $\bm A_p$ is replaced by $\bm A_m$ and the transposed frontal slabs $\bm X_{m,n}^{k^T}$ are replaced by vertical slabs.
\bibliographystyle{plain}
\bibliography{ref,sample-base}

\begin{thebibliography}{10}

\bibitem{auer2007dbpedia}
S{\"o}ren Auer, Christian Bizer, Georgi Kobilarov, Jens Lehmann, Richard
  Cyganiak, and Zachary Ives.
\newblock Dbpedia: A nucleus for a web of open data.
\newblock In {\em The semantic web}, pages 722--735. Springer, 2007.

\bibitem{balazevic2019tucker}
Ivana Balazevic, Carl Allen, and Timothy Hospedales.
\newblock Tucker: Tensor factorization for knowledge graph completion.
\newblock In {\em Proceedings of the 2019 Conference on Empirical Methods in
  Natural Language Processing and the 9th International Joint Conference on
  Natural Language Processing (EMNLP-IJCNLP)}, pages 5188--5197, 2019.

\bibitem{barabasi2016network}
Albert-L{\'a}szl{\'o} Barab{\'a}si et~al.
\newblock {\em Network science}.
\newblock Cambridge university press, 2016.

\bibitem{berlingerio2011foundations}
Michele Berlingerio, Michele Coscia, Fosca Giannotti, Anna Monreale, and Dino
  Pedreschi.
\newblock Foundations of multidimensional network analysis.
\newblock In {\em 2011 international conference on advances in social networks
  analysis and mining}, pages 485--489. IEEE, 2011.

\bibitem{bollacker2008freebase}
Kurt Bollacker, Colin Evans, Praveen Paritosh, Tim Sturge, and Jamie Taylor.
\newblock Freebase: a collaboratively created graph database for structuring
  human knowledge.
\newblock In {\em Proceedings of the 2008 ACM SIGMOD international conference
  on Management of data}, pages 1247--1250, 2008.

\bibitem{bordes2013translating}
Antoine Bordes, Nicolas Usunier, Alberto Garcia-Duran, Jason Weston, and Oksana
  Yakhnenko.
\newblock Translating embeddings for modeling multi-relational data.
\newblock In {\em Advances in neural information processing systems}, pages
  2787--2795, 2013.

\bibitem{bottou2010large}
L{\'e}on Bottou.
\newblock Large-scale machine learning with stochastic gradient descent.
\newblock In {\em Proceedings of COMPSTAT'2010}, pages 177--186. Springer,
  2010.

\bibitem{carlson2010toward}
Andrew Carlson, Justin Betteridge, Bryan Kisiel, Burr Settles, Estevam~R
  Hruschka, and Tom~M Mitchell.
\newblock Toward an architecture for never-ending language learning.
\newblock In {\em Twenty-Fourth AAAI Conference on Artificial Intelligence},
  2010.

\bibitem{chiantini2012generic}
Luca Chiantini and Giorgio Ottaviani.
\newblock On generic identifiability of 3-tensors of small rank.
\newblock {\em SIAM Journal on Matrix Analysis and Applications},
  33(3):1018--1037, 2012.

\bibitem{domanov2013uniqueness}
Ignat Domanov and Lieven De~Lathauwer.
\newblock On the uniqueness of the canonical polyadic decomposition of
  third-order tensors --- part ii: Uniqueness of the overall decomposition.
\newblock {\em SIAM Journal on Matrix Analysis and Applications (SIMAX)},
  34(3):876--903, 2013.

\bibitem{easley2010networks}
David Easley, Jon Kleinberg, et~al.
\newblock {\em Networks, crowds, and markets}, volume~8.
\newblock Cambridge university press Cambridge, 2010.

\bibitem{franz2009triplerank}
Thomas Franz, Antje Schultz, Sergej Sizov, and Steffen Staab.
\newblock Triplerank: Ranking semantic web data by tensor decomposition.
\newblock In {\em International semantic web conference}, pages 213--228.
  Springer, 2009.

\bibitem{harshman1994parafac}
Richard~A Harshman, Margaret~E Lundy, et~al.
\newblock Parafac: Parallel factor analysis.
\newblock {\em Computational Statistics and Data Analysis}, 18(1):39--72, 1994.

\bibitem{himmelstein2015heterogeneous}
Daniel~S Himmelstein and Sergio~E Baranzini.
\newblock Heterogeneous network edge prediction: a data integration approach to
  prioritize disease-associated genes.
\newblock {\em PLoS computational biology}, 11(7), 2015.

\bibitem{himmelstein2017systematic}
Daniel~Scott Himmelstein, Antoine Lizee, Christine Hessler, Leo Brueggeman,
  Sabrina~L Chen, Dexter Hadley, Ari Green, Pouya Khankhanian, and Sergio~E
  Baranzini.
\newblock Systematic integration of biomedical knowledge prioritizes drugs for
  repurposing.
\newblock {\em Elife}, 6:e26726, 2017.

\bibitem{horby2020dexamethasone}
Peter Horby, Wei~Shen Lim, Jonathan~R Emberson, Marion Mafham, Jennifer~L Bell,
  Louise Linsell, Natalie Staplin, Christopher Brightling, Andrew Ustianowski,
  Einas Elmahi, et~al.
\newblock Dexamethasone in hospitalized patients with covid-19-preliminary
  report.
\newblock {\em The New England journal of medicine}, 2020.

\bibitem{drkg2020}
Vassilis~N. Ioannidis, Xiang Song, Saurav Manchanda, Mufei Li, Xiaoqin Pan,
  Da~Zheng, Xia Ning, Xiangxiang Zeng, and George Karypis.
\newblock Drkg - drug repurposing knowledge graph for covid-19.
\newblock \url{https://github.com/gnn4dr/DRKG/}, 2020.

\bibitem{jiang2012link}
Xueyan Jiang, Volker Tresp, Yi~Huang, and Maximilian Nickel.
\newblock Link prediction in multi-relational graphs using additive models.
\newblock {\em SeRSy}, 919:1--12, 2012.

\bibitem{kolda2009tensor}
Tamara~G Kolda and Brett~W Bader.
\newblock Tensor decompositions and applications.
\newblock {\em SIAM review}, 51(3):455--500, 2009.

\bibitem{kolda2005higher}
Tamara~G Kolda, Brett~W Bader, and Joseph~P Kenny.
\newblock Higher-order web link analysis using multilinear algebra.
\newblock In {\em Proceedings of Fifth IEEE International Conference on Data
  Mining}, pages 8--pp. IEEE, 2005.

\bibitem{lin2017learning}
Hailun Lin, Yong Liu, Weiping Wang, Yinliang Yue, and Zheng Lin.
\newblock Learning entity and relation embeddings for knowledge resolution.
\newblock {\em Procedia Computer Science}, 108:345--354, 2017.

\bibitem{newman2018networks}
Mark Newman.
\newblock {\em Networks}.
\newblock Oxford university press, 2018.

\bibitem{nickel2011three}
Maximilian Nickel, Volker Tresp, and Hans-Peter Kriegel.
\newblock A three-way model for collective learning on multi-relational data.
\newblock In {\em Icml}, volume~11, pages 809--816, 2011.

\bibitem{prescott2020corticosteroids}
Hallie~C Prescott and Todd~W Rice.
\newblock Corticosteroids in covid-19 ards: evidence and hope during the
  pandemic.
\newblock {\em Jama}, 324(13):1292--1295, 2020.

\bibitem{rendle2010pairwise}
Steffen Rendle and Lars Schmidt-Thieme.
\newblock Pairwise interaction tensor factorization for personalized tag
  recommendation.
\newblock In {\em Proceedings of the third ACM international conference on Web
  search and data mining}, pages 81--90, 2010.

\bibitem{riedel2013relation}
Sebastian Riedel, Limin Yao, Andrew McCallum, and Benjamin~M Marlin.
\newblock Relation extraction with matrix factorization and universal schemas.
\newblock In {\em Proceedings of the 2013 Conference of the North American
  Chapter of the Association for Computational Linguistics: Human Language
  Technologies}, pages 74--84, 2013.

\bibitem{sanchez1990tensorial}
Eugenio Sanchez and Bruce~R Kowalski.
\newblock Tensorial resolution: a direct trilinear decomposition.
\newblock {\em Journal of Chemometrics}, 4(1):29--45, 1990.

\bibitem{sidiropoulos2017tensor}
Nicholas~D Sidiropoulos, Lieven De~Lathauwer, Xiao Fu, Kejun Huang, Evangelos~E
  Papalexakis, and Christos Faloutsos.
\newblock Tensor decomposition for signal processing and machine learning.
\newblock {\em IEEE Transactions on Signal Processing}, 65(13):3551--3582,
  2017.

\bibitem{singhal2012introducing}
Amit Singhal.
\newblock Introducing the knowledge graph: things, not strings.
\newblock {\em Official google blog}, 16, 2012.

\bibitem{socher2013reasoning}
Richard Socher, Danqi Chen, Christopher~D Manning, and Andrew Ng.
\newblock Reasoning with neural tensor networks for knowledge base completion.
\newblock In {\em Advances in neural information processing systems}, pages
  926--934, 2013.

\bibitem{sorensen2015coupled}
MIKAEL S{\o}rensen, Ignat Domanov, and L~De~Lathauwer.
\newblock Coupled canonical polyadic decompositions and (coupled)
  decompositions in multilinear rank-(lr, n, lr, n, 1) terms—part ii:
  Algorithms.
\newblock {\em SIAM Journal on Matrix Analysis and Applications},
  36:1015--1045, 2015.

\bibitem{suchanek2007yago}
Fabian~M Suchanek, Gjergji Kasneci, and Gerhard Weikum.
\newblock Yago: a core of semantic knowledge.
\newblock In {\em Proceedings of the 16th international conference on World
  Wide Web}, pages 697--706, 2007.

\bibitem{sun2019rotate}
Zhiqing Sun, Zhi-Hong Deng, Jian-Yun Nie, and Jian Tang.
\newblock Rotate: Knowledge graph embedding by relational rotation in complex
  space.
\newblock In {\em International Conference on Learning Representations}, 2018.

\bibitem{yang2014embedding}
Bishan Yang, Wen-tau Yih, Xiaodong He, Jianfeng Gao, and Li~Deng.
\newblock Embedding entities and relations for learning and inference in
  knowledge bases.
\newblock {\em arXiv preprint arXiv:1412.6575}, 2014.

\bibitem{zheng2020dgl}
Da~Zheng, Xiang Song, Chao Ma, Zeyuan Tan, Zihao Ye, Jin Dong, Hao Xiong, Zheng
  Zhang, and George Karypis.
\newblock Dgl-ke: Training knowledge graph embeddings at scale.
\newblock {\em arXiv preprint arXiv:2004.08532}, 2020.

\end{thebibliography}
\end{document}